\documentclass[multphys,vecphys]{svmult}

\usepackage{makeidx}         % allows index generation
\usepackage{graphicx}        % standard LaTeX graphics tool
\usepackage{multicol}        % used for the two-column index
\usepackage[bottom]{footmisc}% places footnotes at page bottom
\makeindex             % used for the subject index
\begin{document}

\title*{Previously unobserved water lines detected in the post-impact spectrum}
\author{R.~J.~Barber, S.~Miller, T.S.~Stallard, J.~Tennyson}
\institute{University College London, Gower Street, London WC1E 6BT}
% the following line commented out to prevent a first page with my email
%\texttt{bob@theory.phys.ucl.ac.uk}

\maketitle

%%%%%%%%%%%%%%% here goes the author index entry   %%%%%%%%%%%%%%%%%%%%%% 
% 
%  \index{C}{Authorfamilyname1, A.C.}
%  \index{C}{Authorfamilyname2, B.A.}
%
%%%%%%%%%%%%%%%%%%%%%%%%%%%%%%%%%%%%%%%%%%%%%%%%%%%%%%%%%%%%%%%%%%%%%%%%%

\setcounter{footnote}{0}

%
% abstract is optional; as an example I have added an old one of mine
%

\section{Water lines in Tempel 1}

The team from UCL, monitored the Deep Impact event using the CGS4
spectrometer on the United Kingdom Infrared Telescope, UKIRT. Our
principal objective was to determine the temporal development of solar
pumped fluorescent (SPF) transitions of H$_2$O following impact, and
to interpret the results using the recently published \emph{ab inito}
water line list, BT2 (Barber et al., 2006a). BT2 was produced at UCL
using the DVR3D suite of programs (Tennyson et al., 2004), and is the
most complete and most accurate water line list in existence.

Normal observing techniques were employed: these and other aspects of
our work are fully reported in Barber et al.~(2006b). We obtained
spectra in the wavelength range centred on 2.894 $\mu$m, with a
spectral range of $\pm$0.040 $\mu$m. Apart from containing a number of
SPF transitions (Dello Russo et al., 2004), it is largely devoid of
other molecular lines (such as CO), which makes it possible to model
the region using the BT2 line list without having to include other
species. However, in the subsequent examination of the data, in order
to maximise the S/N ratio, we restricted our analysis to a narrower
wavelength range: 2.8945 - 2.8985 $\mu$m.

It had been our intention to obtain spectra of Tempel 1 on the night
prior to impact, on impact night, and on the night after
impact. However, prior to impact the comet was not sufficiently bright
for us to be able to obtain useful data, and on the night after
impact, reduced intensity and a deterioration in the observing
conditions prevented us obtaining high quality data. Our results
therefore relate only to the period of 143 minutes immediately
following impact.

Our signal was effectively confined to one pixel row (we estimate that
this contained $\sim$65\% of the signal), and it attests to the
accuracy of the UKIRT tracking system that there was no detectable
drift from this position over the whole observing session. No attempt
was made to recover the small amount of signal from adjacent rows, as
an analysis of the data revealed that this would have resulted in a
reduction in the overall S/N ratio. Moreover, because of the weakness
of the signal, it was necessary to combine the data for the whole
observing run in order to obtain a useful S/N ratio, which meant that,
taking the spectral region as a whole, no temporal resolution could be
obtained (see below for comments on temporal resolution for specific
groups of lines).

\begin{figure}[ht]
\begin{center}
\includegraphics[angle=270,width=1.0\textwidth]{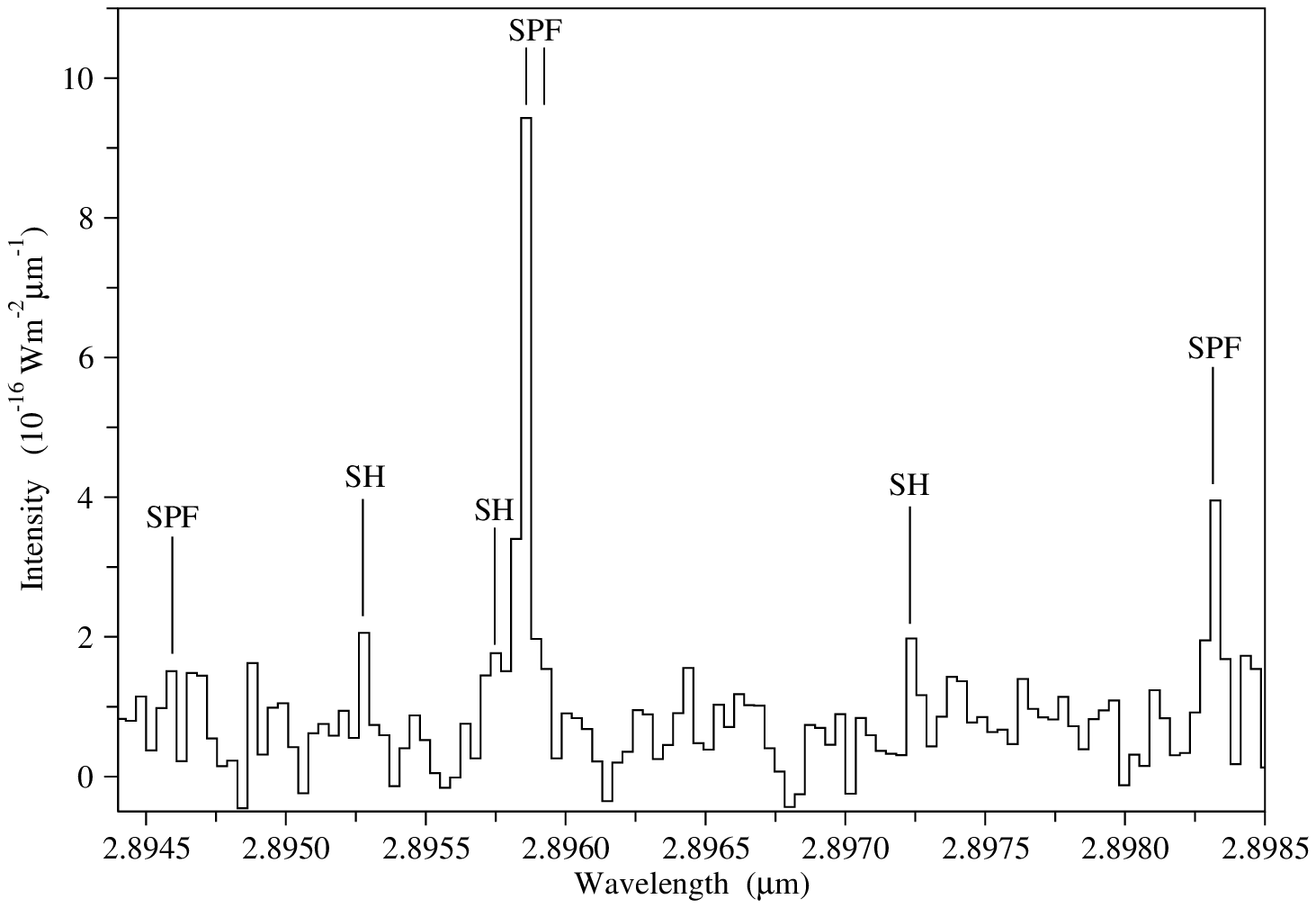}
\caption{Observed post-impact spectrum Tempel 1. Wavelength is in the
rest frame. Reproduced from Barber et al.~(2006b).
\label{Tempel1observed}}
\end{center}
\end{figure}

The spectrum for the period July 4 05:54 to 08:17 U.T. is shown in
Figure \ref{Tempel1observed}. The fact that some of the intensities
are negative may in part be due to our having over-corrected for the
continuum, or more likely, is due to noise, which we estimate to be in
the region of 0.4x10$^{-16}$ Wm$^{-2}$ $\mu$m$^{-1}$.

Two bright fluorescent transitions at 2.89580 and 2.89831 cm$^{-1}$,
stand out against a background of weaker features.  Many of these
weaker features are close to, or below, the noise threshold and these
we disregard. However, there are a several where we estimate the S/N
ratio to be greater than 4 and which, unlike noise, appear in the same
place (albeit with differing intensities) in many of the individual
frames. We interpret these as also being genuine signals and have
marked them either SPF (where their wavelengths correspond to known
fluorescent transitions), or SH, (in those cases where transitions are
thought to be by some other route). In the first column of Table
\ref{Temp1assign} we give the observed wavelength of each of the
features (adjusted for red shift).

The position of many of the spectral features in Figure
\ref{Tempel1observed} can be replicated using the BT2 synthetic water
line list. Some of these are identified as being due to SPF
transitions (sometimes blended). However, some of the features did not
correspond to known SPF transitions, and moreover, as far as we are
aware, have not previously been recorded in cometary spectra. We have
labelled these `SH' (the acronym is convenient as it can stand for
`solar heating', which is definitely true, or `stochastic heating',
which may also be an apt description). It should be noted that
although the line positions derived from BT2 agree well with those of
the observed features, the intensities are only approximate guides, as
the LTE assumption used by BT2 is not generally valid for cometary
spectra. Moreover, since the mechanism behind the formation of the SH
transitions is not known, a precise match between the intensity of the
observed features and the BT2 synthetic spectra is not to be expected.

\begin{table}
\begin{center}
\caption{Assignments of SPF and previously unobserved SH lines in the
post-impact spectrum of Tempel 1. The first
column gives the observed wavelength of
each of the features (adjusted for red shift), the next column
identifies the transition: the vibrational quantum numbers are given
in round brackets and the rotational quantum numbers in square
brackets. The last three columns give: the
experimentally-determined wavelength of the listed transition, the
Einstein A coefficient computed using BT2, and our designation of type:
SPF or SH. Reproduced from Barber et al.~(2006b).
\label{Temp1assign}}
\smallskip
\begin{tabular}{cccrc}
\hline
$\lambda_{observed}$ & Identification & $\lambda_{laboratory}$ & A$_{if}$ & Type \\
$\mu$m $\pm$0.00005 & (see text for notation) & $\mu$m & s$^{-1}$ &   \\[0.5ex]
\hline		
2.89458	& (101)[211] - (001)[220] & 	2.89462 &  1.9 & SPF \\
2.89527	& (103)[110] - (102)[110] &	2.89526	& 53.5 &  SH \\
2.89527	& (211)[322] - (210)[211] &	2.89528	&  8.5 &  SH \\
2.89573	& (210)[101] - (011)[000] &	2.89570	&  5.1 &  SH \\
2.89580	& (200)[110] - (100)[221] &	2.89578	&  4.7 & SPF \\
2.89591	& (101)[202] - (100)[321] &	2.89590	&  1.7 & SPF \\
2.89723	& (220)[212] - (021)[111] &	2.89728	&  4.4 &  SH \\
2.89831	& (200)[110] - (001)[111] &	2.89830	&  6.6 & SPF \\
\hline
\end{tabular}
\end{center}
\end{table}

\begin{figure}
\begin{center}
\includegraphics[angle=270,width=1.0\textwidth]{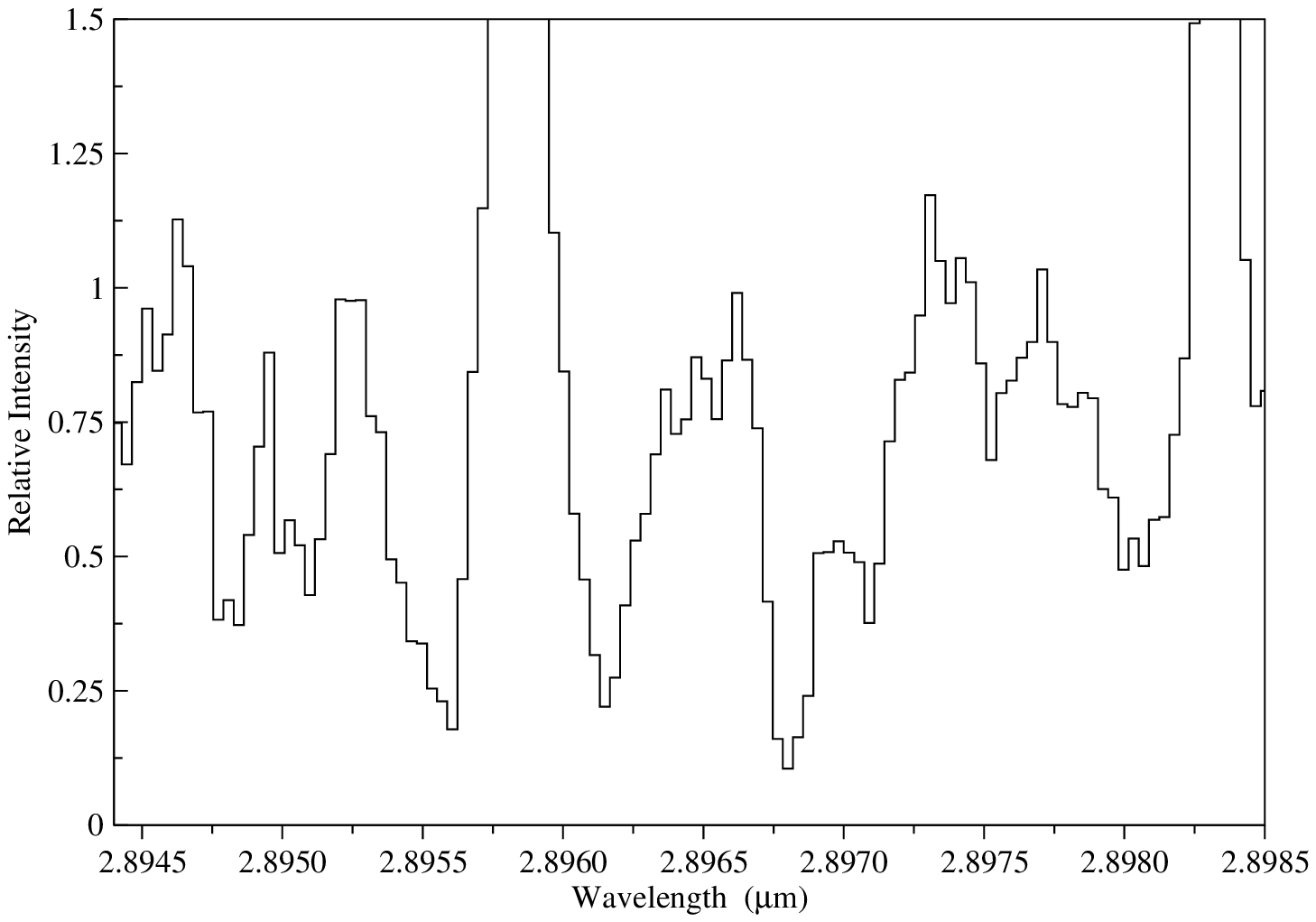}
\caption{Observed post-impact spectrum Tempel 1, moving average of 5
pixels. Wavelength is in the rest frame. The vertical scale has been
limited to assist identification of the weaker features. Reproduced
from Barber et al.~(2006b).
\label{Tempel1obsmvgav}}
\end{center}
\end{figure}

\begin{figure}[ht]
\begin{center}
\includegraphics[angle=270,width=1.0\textwidth]{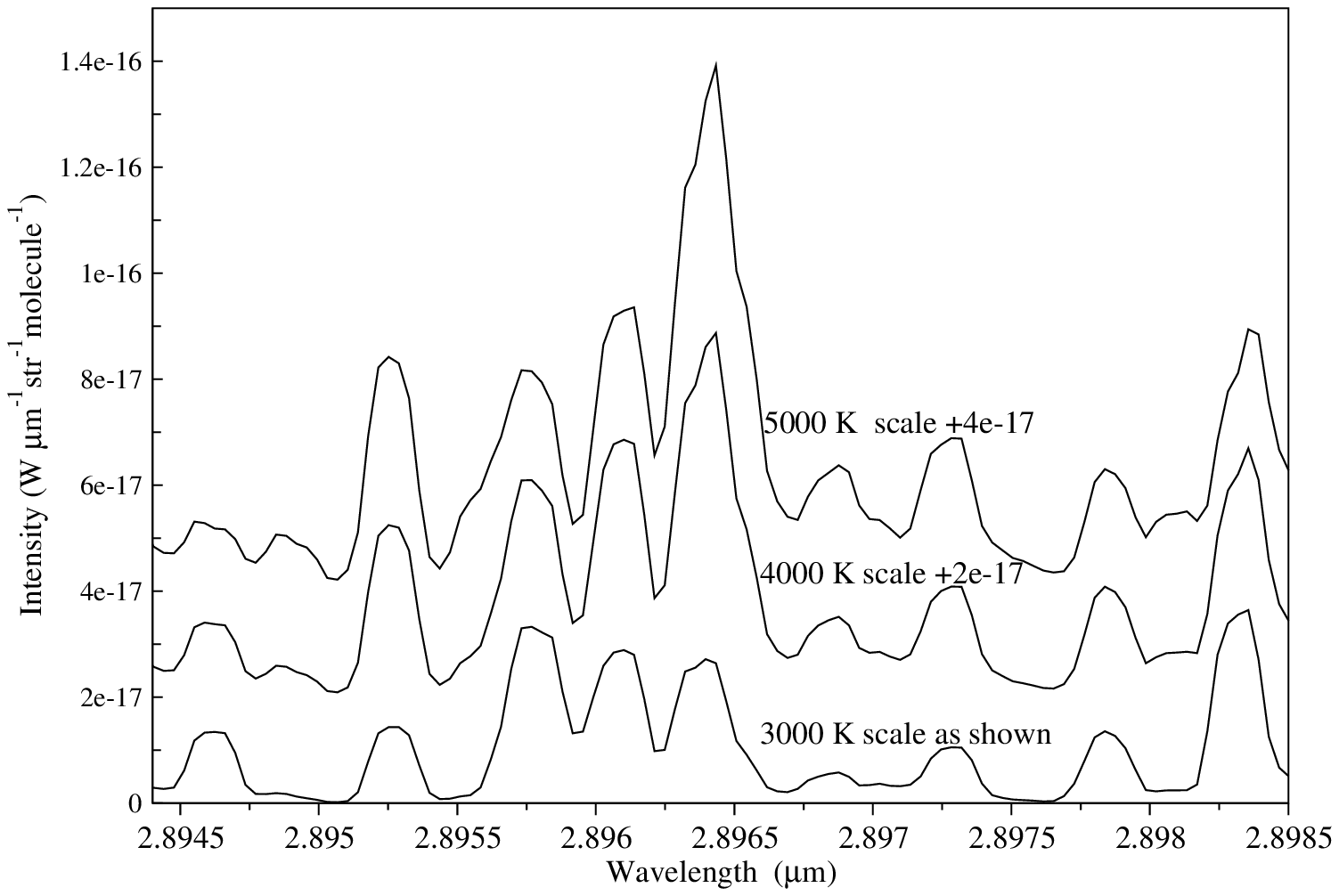}
\caption{BT2 synthetic spectra at 3\,000 K, 4\,000 K and 5\,000 K,
J$_{max}$=3.  Reproduced from Barber et al.~(2006b).
\label{Tempel1synthJ3}}
\end{center}
\end{figure}

\begin{figure}[ht]
\begin{center}
\includegraphics[angle=270,width=1.0\textwidth]{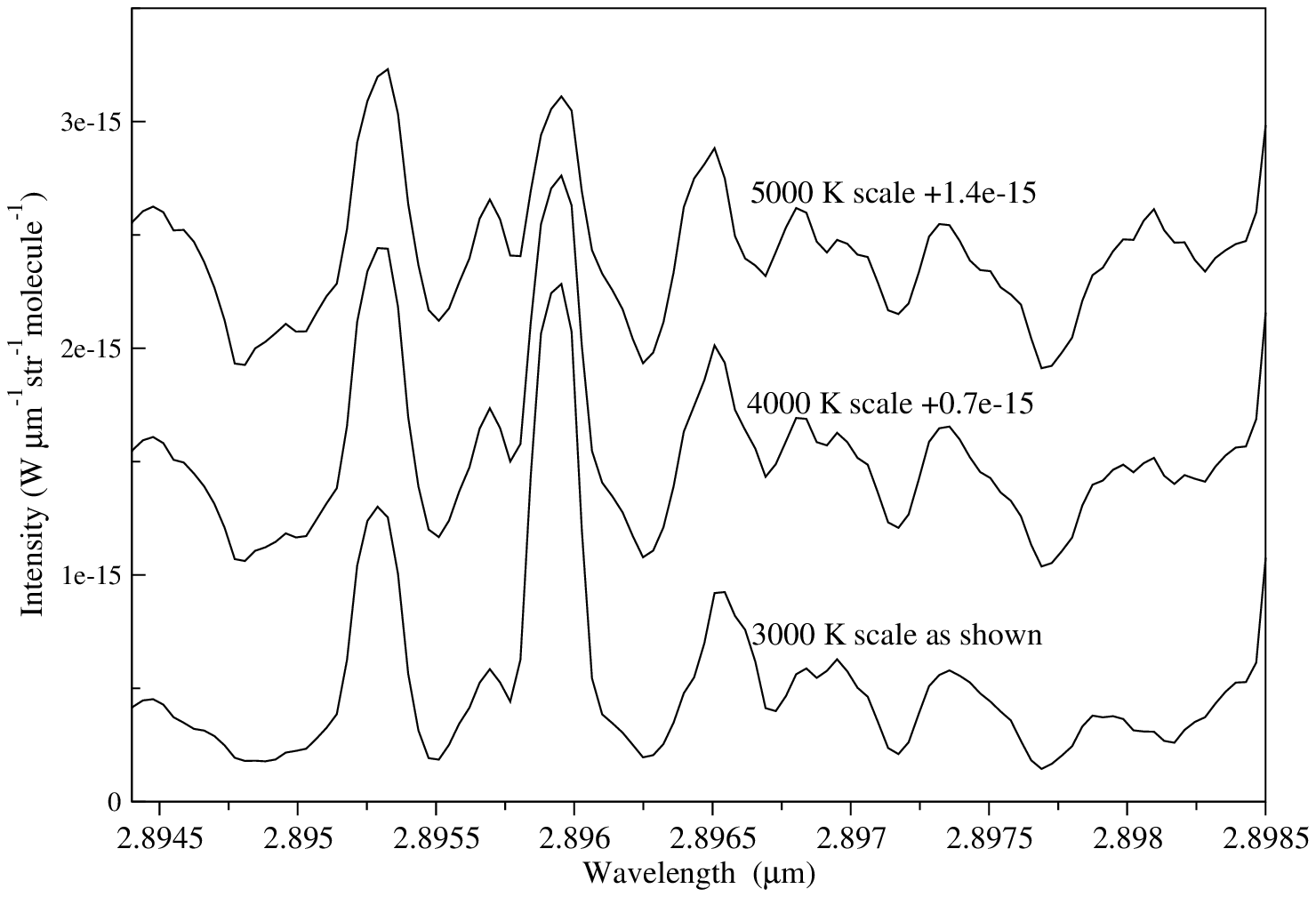}
\caption{BT2 synthetic spectra at 3\,000 K, 4\,000 K and 5\,000 K,
J$_{max}$=50.  Reproduced from Barber et al.~(2006b).
\label{Tempel1synthJ50}}
\end{center}
\end{figure}

In attempting to identify the non-SPF features in our observed
spectrum, we generated a series of synthetic spectra using BT2, some
of which are shown in Figures \ref{Tempel1synthJ3} and
\ref{Tempel1synthJ50}. In order to improve the signal to noise ratio
and also to lessen the significance of errors in the wavelength
calibration of our detector, estimated to be in the region of 0.00004
$\mu$m (slightly greater than a single pixel width, which is 0.000033
$\mu$m), we artificially reduced the resolution of our observed data
from an instrument-limited $\sim$37\,000 to a pixel-averaged limited
value of $\sim$17\,500 by taking a moving average of five pixels in the
wavelength dimension in order to assist in matching the observed and
synthetic data.  Figure \ref{Tempel1obsmvgav} shows our observed
spectrum with this artificially de-graded spectral resolution. In this
figure the vertical scale is terminated at a level well below the peak
intensities of the two strongest SPF transitions. This was done to aid
identification of the weak transitions.

Figure \ref{Tempel1synthJ3} shows three synthetic spectra generated at
3\,000, 4\,000 and 5\,000 K (assuming LTE), using the BT2 line list
(with some corrections to the calculated wavelengths based on the
available experimental data).  In producing these spectra, we applied
the restriction that only states with J$\le$3 are included. The
resolution in Figure \ref{Tempel1synthJ3} was set to be the same as
Figure \ref{Tempel1obsmvgav}. Figure \ref{Tempel1synthJ50} shows
another set of synthetic BT2 spectra generated at the same three
temperatures, but this time including all J levels (up to 50). Some of
the same features are observed as in Figure \ref{Tempel1obsmvgav}, but
we note that Figure \ref{Tempel1synthJ3} matches the observed spectrum
better than Figure \ref{Tempel1synthJ50} does.

Figure \ref{Tempel1synthJ3} is therefore a high-v, low-J spectrum and
it reproduces well the position (less-so, the intensities) of many of
the non-SPF features in the low-resolution observed spectrum, Figure
\ref{Tempel1obsmvgav}. It should be noted that similar features are
observed in synthetic spectra generated for temperatures greater than
3\,000 K and the appearances of the spectra vary little once T$_{vib}$
$>$ 4\,500 K, except for differences in the general levels of
intensity. It is also noted that there are other features in our
observed spectrum Figure \ref{Tempel1obsmvgav} that are replicated in
the synthetic spectrum Figure \ref{Tempel1synthJ3}. These are also due
to water emission and are produced by the blending of many overlapping
transitions from different vibrational manifolds. They have not been
included in Table \ref{Temp1assign} as it is not possible to assign
them to one or two individual transitions.

\subsection{Assigning the features}
\label{Temp1feat}
Among the assigned features are fluorescent emission
lines from levels that have two quanta of H$_2$O stretching, such as
(2 0 0)$\rightarrow$(1 0 0) and (1 0 1)$\rightarrow$(1 0 0): in the
former transition, the emission involves one quantum of $\nu_1$, in
the latter it involves one quantum of $\nu_3$. A quantum of $\nu_1$
and of $\nu_3$, have similar energies $\sim$3\,450 cm$^{-1}$. In both
cases, the final state is the 1$\nu_1$ state. Because a quantum of
$\nu_2$ carries less than half the energy of a $\nu_1$ or
$\nu_3$ quanta, ($\sim$1\,500 cm$^{-1}$), transitions involving a
change of one quantum of $\nu_2$ are not observed in our selected
wavelength range.
 
Our post-impact spectrum of Tempel 1 also includes several transitions
from states that include one or more quanta of $\nu_2$, or that
involve a total of 4 vibrational quanta. These are not SPF spectral
features.  It will be seen from Table \ref{Temp1assign} that these
include the blend of (1 0 3)$\rightarrow$(1 0 2) and (2 1
1)$\rightarrow$(2 1 0) at 2.8953 $\mu$m, and (2 2 0)$\rightarrow$(0 2
1) at 2.8972 $\mu$m. It seems likely that upper levels with more than
two vibrational quanta have not been populated by solar pumping from
ground vibrational states, but by another mechanism. Whereas SPF
transitions originate from upper states having energies in the region
of 7\,300 cm$^{-1}$, the transitions that we have labelled `SH' all
originate from higher energy states (those upper states having four
vibrational quanta are in the energy range 10\,300-14\,400 cm$^{-1}$.
It is possible that the
production route involves H$_2$O molecules that have sublimed from the
freshly exposed icy grains ejected by the impact. However, an
understanding of the precise nature of these SH lines will require
further research. We have recently learned that Villanueva et
al.~(2006) observed some previously unrecorded lines in the 2.8313
$\mu$m region in the post-impact spectrum of Tempel 1, which may be
due to transitions from higher energy ro-vibrational states of
H$_2$O.

\subsection{Comparison with other spectra} 
We have compared our results with the
spectra obtained by Mumma et al.~(2005). 

\begin{figure}[ht]
\begin{center}
\includegraphics[angle=270,width=1.0\textwidth]{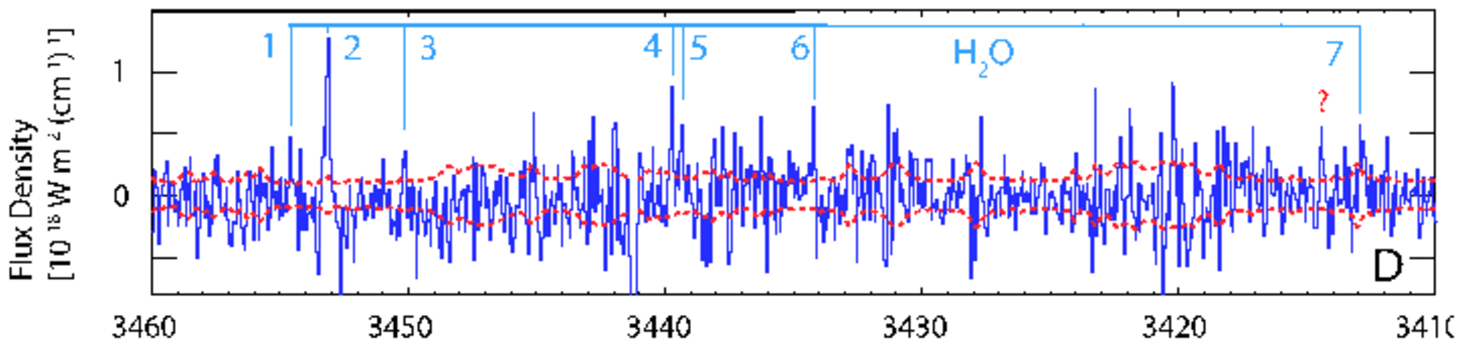}
\caption{Spectrum of Tempel 1 post-impact (abscissa is frequency in
units of cm$^{-1}$). Reproduced from Mumma et al.~(2005) - spectrum D.
\label{MummachartD}}
\end{center}
\end{figure}

As far as the positions of the observed SPF transitions are concerned,
our results agree well with those of Mumma et al.. However, none of
the SH features that we identify appear in Mumma et al.'s spectrum
D. One possible reason for the difference between our results and
those of Mumma et al.~is the difference in times when the spectra were
obtained.  Our results were obtained between 05:54 and 08:17 U.T. on
impact night, whilst Mumma et al.'s spectrum in Figure
\ref{MummachartD} were obtained between 6:43 and 7:25 U.T.. By summing
the intensities of all the SH features, we were able to achieve a
degree of temporal resolution that was not possible for the individual
lines. We observed that the total intensity of the SH features was
particularly strong during the period 20--40 minutes after impact.,
but by 50 minutes after impact had declined to a level that was only
slightly above that of the background noise. This could be the reason
why the features are not observed in Mumma et al.'s spectra..

Recently we have obtained UKIRT spectra of Comet
73P/Schwassmann-Wachmann, fragment-C in order to investigate whether
this recently fragmented comet exhibits similar SH features to those
that we observed in Temple 1. Our preliminary investigation of the 73P
spectra does not show any of the SH features. Work on this is continuing.

\chapter*{Bibliography}
\begin{itemize}
\item[]
Barber R.J., Tennyson J., Harris G.J., Tolchenov R., 2006a, MNRAS, 368, 1087
\smallskip
\item[] 
Barber R.J., Miller S., Stallard T., Tennyson J. et al., 2006b
`UKIRT Deep Impact observations: light curve, ejecta expansion rates
and water spectral features', Icarus (submitted).
\smallskip
\item[] 
Dello Russo N.,DiSanti M.A., Magee-Sauer K., Gibb E.L., Mumma
M.J., Barber R.J., Tennyson J., 2004, Icarus, 168, 186
\smallskip
\item[] 
Mumma M.J., Disanti M.A., Magee-Saurer K., Bonev B.P.,
Villanueva G.L., et al., 2005, Science, 310, 270
\smallskip
\item[] 
Tennyson J., Kostin M., Barletta P., Harris G.J., Polyansky
O.L., Ramanlal J., Zobov N.F., 2004, Comp.~Phys.~Comm., 163, 85
\smallskip
\item[] 
Villanueva G.L., Bonev B.P., Mumma M.J., DiSanti M.A.,
Magee-Sauer K., 2006, `Infrared spectral survey of the ejecta of comet
Tempel 1 using NIRSPEC/Keck-2' Springer-Verlag ESO Astrophysical
Symposia Series (Deep Impact as a World Observatory Event)
ed. K\"{a}ufl U., Sterken C.

\end{itemize}

%%%%%%%%%%%%%%%%%%%%%%%%%%%%%%%%%%%%%%%%%%%%%%%%%%%%%%%%%%%%%%%%%%%%%%  }

%%%%%%%%%%%%%%%%%%%%%%%%%%%%%%%%%%%%%%%%%%%%%%%%%%%%%%%%%%%%%%%%%%%%%%

\printindex
\end{document}